# Topological Photonic Quasicrystals: Fractal Topological Spectrum and Protected Transport

Miguel A. Bandres,[1] Mikael C. Rechtsman,[1,2] and Mordechai Segev[1]

[1]*Physics Department and Solid State Institute, Technion, 32000 Haifa, Israel*
[2]*Department of Physics, The Pennsylvania State University, University Park, Pennsylvania 16802, USA*

We show that it is possible to have a topological phase in two-dimensional quasicrystals without any magnetic field applied, but instead introducing an artificial gauge field via dynamic modulation. This topological quasicrystal exhibits scatter-free unidirectional edge states that are extended along the system's perimeter, contrary to the states of an ordinary quasicrystal system, which are characterized by power-law decay. We find that the spectrum of this Floquet topological quasicrystal exhibits a rich fractal (self-similar) structure of topological "minigaps," manifesting an entirely new phenomenon: fractal topological systems. These topological minigaps form only when the system size is sufficiently large because their gapless edge states penetrate deep into the bulk. Hence, the topological structure emerges as a function of the system size, contrary to periodic systems where the topological phase can be completely characterized by the unit cell. We demonstrate the existence of this topological phase both by using a topological index (Bott index) and by studying the unidirectional transport of the gapless edge states and its robustness in the presence of defects. Our specific model is a Penrose lattice of helical optical waveguides—a photonic Floquet quasicrystal; however, we expect this new topological quasicrystal phase to be universal.

Topological insulators are materials that insulate in their bulk but allow extremely robust flow of current on their surfaces [1–4]. In two dimensions, they possess one-way edge currents that are immune to scattering and imperfections since their physical properties are related to topological quantities rather than geometric ones. This robust transport, which was first manifested in the highly robust quantum Hall conductivity measurements [5], has potential applications in novel materials for topological quantum computing [6,7] and robust spintronic devices [8]. The concept of topological protection is not limited to electronic systems; rather, it is universal: it was recently demonstrated for photons [9–11], cold atoms [12,13], mechanical systems [14,15], and was also predicted for exciton polaritons [16], acoustics [17], and phonon systems [18]. Fundamentally, topological protection of photons is profoundly different than for electrons, because photons do not respond to magnetic fields and do not exhibit the spin-orbit interaction that, together with Kramers degeneracy, can provide topological protection in fermionic systems. Consequently, photonic topological insulators [19] have to rely on mechanisms other than those that support their electronic counterparts [22]. Chronologically, the photonic equivalent of the quantum Hall effect was predicted [9,23] and observed first among the photonic topological systems in 2D [9] (using a similar technique, topological effects were also observed in 3D systems [24,25]). The underlying mechanism was the magnetic response of gyro-optic materials, and the experiments took place in the microwave regime, where the magnetic response is relatively strong. However, such magneto-optic effects are very weak at optical frequencies, which prompted a major theoretical effort to realize topological insulation and protection in the optical frequency range [26–29]. This culminated in the first experimental observation of photonic topological insulators in optics [10], as demonstrated in a honeycomb array of evanescently coupled helical waveguides. This first experiment relied on a new concept known as "Floquet topological insulators" [30–33], where the topological features result from periodic temporal modulation, rather than from applied magnetic field [5,9,23] or spin-orbit interaction [1]. Another kind of photonic topological insulator was demonstrated in a silicon-chip platform using a lattice of coupled resonators [11], (see also discussion in [34]). It is already clear that introducing topological protection into photonics [22] enables a host of new applications, for example, robust optical delay lines [27] or novel designs for optical isolators [29]. Also, and more important from the fundamental point of view, photonics also provides an entirely new platform for probing novel topological phenomena that would be extremely difficult or impossible to access in the solid state [35].

In general, physical systems may be classified according to their wave functions: extended, exponentially localized,

and critical (power-law decay). These three types of wave functions uniquely correspond to absolutely continuous, discrete (pointlike), and singular continuous spectra, respectively. For example, for a periodic system all states are extended due to the Bloch's theorem. On the other hand, for a 1D or 2D disordered system all states are exponentially localized due to Anderson localization. Thus far, topological insulators with zero magnetic field have been found in periodic structures, with continuum spectra [10,30,37], and were predicted to also occur in periodic systems containing disorder, where the spectrum is discrete [36,38]. Therefore, an intriguing question to ask is whether it is possible to induce a topological phase in a quasicrystal (QC), even though its eigenstates and its transport properties are entirely different from those associated with periodic crystalline systems and also from the Anderson states characteristic of disordered systems. In this context, QCs are unique structures with long-range order but no periodicity whatsoever [39]. They constitute an intermediate phase between fully periodic lattices and fully disordered media. They lack translational symmetry, as in disordered materials; nevertheless, they possess long-range order and display Bragg diffraction spot as crystals do, but with noncrystallographic rotational symmetry. Because of these characteristics, the fundamental physical features of QCs have unusual properties [40], such as phason fluctuations [41] and extremely low friction [42]. Interestingly, the question of how waves (electrons, photons, or otherwise) are transmitted through a QC is not fully resolved to this day. Its complexity stems from the fact that the lack of periodicity excludes the possibility of describing QCs with the well-known analytic tools such as Bloch's theorem. Indeed, transport in QCs is extremely rich, which is evidenced, for example, by the observation of several phenomena: a surprising regime of disorder-enhanced transport in a photonic 2D QC [43], the localization phase transition of the Aubry-Andre model in 1D quasiperiodic photonic lattices [44], and the fractal energy spectrum of a polariton gas in a 1D Fibonacci quasiperiodic potential [45,46]. The first connection between topology and quasiperiodicity [47] was done by showing that the Aubry-Andre model in 1D quasiperiodic lattices can be exactly mapped to 2D quantum Hall system on a square lattice. The topological properties of these 1D quasiperiodic models, with 0D edge states, were experimentally demonstrated by adiabatic pumping [47] and by bulk phase transitions [48] in a 1D photonic quasicrystal [49]. However, fundamentally, 1D systems cannot exhibit transport along the edges; hence, topologically protected transport cannot be studied in 1D systems. Recently, 2D quasicrystals were studied in the presence of a uniform magnetic field [51], and were found to display a Hofstadter-butterfly-like structure and chiral edge states. However, to date, any connection between topological insulators and the fractal self-similar spectrum structure of the QC has never been suggested nor has it ever been explored.

In principle, a topological phase with zero magnetic field is not expected in a two-dimensional or a three-dimensional QC. This is the case because topological insulators must have at least one extended edge state, scaling with the perimeter of the system, whereas the eigenstates of QCs are generally localized, exhibiting a power-law self-similar (fractal-like) decay [52]. Moreover, the topological edge states of 2D topological insulators are propagating (nonzero velocity) one-way states, while the eigenstates of QCs are generally stationary (zero velocity, like the "immobile" Anderson states in a random potential). Finally, even if some mechanism (such as spin-orbit interaction or temporal modulation) is able to transform some lattices into being topological, there is no guarantee that the same mechanisms would be able to make the quasicrystal topological.

Here, we show that 2D quasicrystals have a topological insulating phase that exhibits mobile one-way edge states residing in the topological gap. These topologically protected (scatter-free) edge states are extended along the system's perimeter (due to their confinement to the edges), contrary to the states of a general QC system, which have power-law decay [52]. Furthermore, the states are induced with zero magnetic field (in a manner similar to the quantum anomalous Hall effect [37]), but by breaking time-reversal symmetry using an artificial gauge field with zero magnetic component. Importantly, we show that the topological phase supports a rich structure of topological "minigaps" whose gapless edge states penetrate deep into the bulk. As such, the topological minigaps form only when the system size is sufficiently large. This is an entirely new phenomenon that endows the "fractal-like" (singular continuous) energy spectrum of QCs with topological properties. In this way, the QC topological structure evolves as a function of the system size, in contrast with periodic systems where the topological phase can be completely characterized using its unit cell. We demonstrate the existence of this topological phase both by using a discrete topological index, i.e., the Bott index [53], and also by studying the topologically protected dynamics (robust unidirectional transport) of the gapless edge states. Finally, although here we specifically study a Penrose lattice [54,55] of helical optical waveguides, we expect this new topological quasicrystal phase to be universally applicable to a variety of quasicrystalline structural variants, including those with higher-order noncrystallographic rotational symmetries.

We begin by describing the physical system, which we take as an experimentally realizable example: a 2D lattice of waveguides that are coupled evanescently—a 2D photonic lattice [56–58]. Two-dimensional photonic QC lattices have been demonstrated in the past in both the optical [59] and the microwave regimes [60], but they were all nontopological. The propagation of optical waves in photonic lattices can be adequately described by the paraxial wave equation, which is mathematically equivalent to the Schrödinger equation—with the propagation coordinate $z$ playing the role of time, and the local change in the refractive index being the potential [61]. Hence, the propagation of waves in this lattice is equivalent to the temporal evolution of a quantum particle in the lattice. Some of the paraxial photonic lattices can be made topological by introducing periodic modulation in $z$, giving rise to an artificial gauge field, in the spirit of Floquet topological insulators [10,30,33]. However, many modulated lattices (for example, all the lattices with a single atom per unit cell) do not become topological, so whether or not a QC can be made topological via modulation is far from being straightforward.

The first experimental photonic Floquet topological insulator [10] was based on such a system, employing a honeycomb lattice of helical waveguides. That particular photonic topological insulator was implemented in "photonic graphene," an array of waveguides arranged as a honeycomb lattice, where the modulation that makes it topological was implemented my making the waveguides helical. Here, we are interested in studying whether a quasicrystal lattice of helical waveguides develops a topological phase.

In the reference frame in which the helical waveguides are stationary, the system can be described by a tight-binding Hamiltonian subject to a vector potential [10] given by

$$i\partial_z \psi_n = \sum_{\langle m \rangle} c_{mn} e^{iA(z) \cdot r_{mn}} \psi_m, \quad (1)$$

where the summation is taken over neighboring lattice sites, $\psi_n$ is the amplitude in the $n$th site, $c_{mn}$ is the coupling constant between site $m$ and $n$, $r_{mn}$ is the displacement between sites $m$ and $n$ (normalized to the lattice constant), $A(z) = A_0[\cos(\Omega z), \sin(\Omega z)]$ is the vector potential (reflecting the helicity of the waveguides) [10], $A_0$ is a measure of the spinning radius, and $\Omega$ is the longitudinal frequency of the helices. We emphasize that our system here is a QC, not a periodic lattice; hence, the problem is considerably more complicated.

Because the Hamiltonian [Eq. (1)] is $z$ dependent, there are no static eigenmodes. However, the Hamiltonian is $Z = 2\pi/\Omega$ periodic; hence, we can use the Floquet theory [30]. That is, the solutions are described as Floquet eigenmodes of the form $\psi_n(z) = \exp(i\beta z)\varphi_n(z)$, where $\varphi_n(z)$ is $z$ periodic and $\beta$ is the Floquet eigenvalue or "quasienergy," which are defined modulo the frequency $\Omega$. Specifically, we can define the time-independent effective Floquet Hamiltonian $H_F$, whose eigenvalues are the quasienergies $\beta$ of our system, and it is given by

$$e^{-iH_F Z} = P \exp\left(-i \int_0^Z H(z) dz\right), \quad (2)$$

where $P$ is the path-ordered exponential operator.

To construct the QC, we arrange the lattice sites in a Penrose tiling [54,55], consisting of two rhombuses (fat and thin) that tile the 2D plane completely in an aperiodic way. The Penrose tiling can be constructed through local matching rules [54] or by de Bruijn's "cut and project method," where a "cut" slice of a five-dimensional (simple cubic) lattice is projected onto the 2D plane [62]. There exist infinitely many noncongruent realizations of a Penrose lattice, which cannot be differentiated by their physical properties because their local environment is the same for all.

In our setting, we place a helical waveguide in each vertex of a Penrose tiling to form a photonic quasicrystal, as depicted in Fig. 1(c). Since the waveguides are coupled evanescently, the coupling decreases exponentially with the distance between waveguides $d$. As such, we model the tight-binding coupling through $c_d = C_0 \exp(-d/d_0)$. Because the coupling is exponentially decreasing, we just consider the first three nearest neighbors, as depicted in Fig. 1(d); their separation distances are $d = \{d0, d1, d2\} = \{1/\tau, 1, \sqrt{3-\tau}\}a$, where $a$ is the size of the edges of the

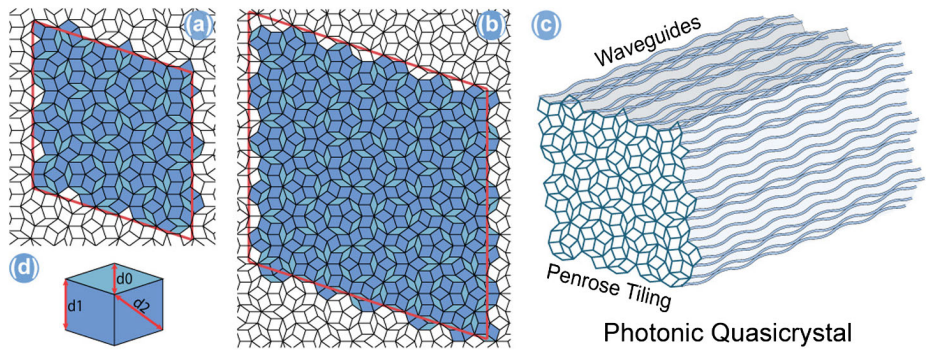

FIG. 1. (a),(b) Periodic Penrose approximants of Penrose quasicrystals, containing 199 and 521 vertices. The red perimeter line defines the unit cell, which can be replicated periodically to fill a 2D plane, or—as we do here—folded to form a torus. (c) Sketch of the photonic Floquet quasicrystal formed by introducing a helical waveguide in each vertex of a Penrose tiling. (d) Nearest-neighbor distances $d_0 < d_1 < d_2$ used in the tight-binding model describing the quasicrystal.

Penrose tiling (it represents the second nearest neighbor) and $\tau = (1+\sqrt{5})/2$ is the golden ratio.

Since the topological insulating phase is characterized by gapless edge states (i.e., edge states that exist at all energies within a band gap), it is important to correctly identify the gaps in the spectrum of our system. Because the QC lacks periodicity, we take the finite system approach where the notion of a gap cannot be defined for a single system but as a property of a sequence of systems in the large size limit [63]. Also, a gapped system (not necessarily topological) can have in-gap states localized at the spatial boundary. To avoid such nontopological states, it is convenient to impose periodic boundary conditions [64]; i.e., the two-dimensional lattice lives in a torus. Of course, in a QC it is not possible to impose periodic boundary conditions due to its lack of periodicity. It is, however, possible to generate a sequence of periodic lattices with growing unit cells that approximate the infinite QC in a systematic way [65,66]. These periodic lattices are called QC tiling approximants. Here, we use the periodic Penrose tiling approximant [66]. These approximants form a series of periodic lattices with a unit cell in the shape of a thick Penrose rhombus and increasing size. That is, the number of vertices is $N = 29, 76, 199, 521, 1364, 3571, 9349, 24\,476, \ldots$. As an example, Figs. 1(a) and 1(b) depict two approximants and their unit cell. The approximants we use display an important property: they have the minimum number of defects (two per units cell) irrespective of the unit-cell size, while for other approximants the number of defects increases as the unit cell is increased (here, a defect is an edge where the Penrose matching rules are violated) [67]. Therefore, a Penrose approximant reproduces the same local configuration as the original Penrose tiling, except at the two "defects," as far as the unit cell is concerned. These two defects become negligible as the size of the approximant increases.

Before studying the topological properties of our photonic QC, we first discuss its quasicrystalline features. To get a physical insight into the properties of a QC, it is important to understand its two main features: aperiodicity and quasiperiodic order. Aperiodicity simply means there is no translation that maps the quasicrystal lattice onto itself. The quasiperiodic order, or "long-range" order, refers to the repetitivity of the lattice (self-similarity). That is, any finite-size region reappears again and again in a non-periodic fashion throughout the QC lattice, but always having different surroundings. This quasiperiodic order is described mathematically by a property known as Conway's theorem [55], which states that in a QC a given local pattern in a circular region of diameter $d$ is never more than $2d$ away from an exactly identical region. The existence of critical states with power-law decay in a QC is intimately related to the competition between these two properties [66]. The absence of periodicity acts similarly to random disorder, and favors the localization of the wave function in a local pattern. However, due to Conway's theorem, this local pattern must have duplicates that extend throughout the whole lattice. Then, a hopping mechanism causes resonances between such localized states at identical local configurations. This makes the wave function more extended, and, consequently, the wave function exhibits power-law decay and a self-similar structure, in the sense that portions of the wave function over identical configurations differ only by a scale factor. In this way, each bulk band is characterized by a local pattern localized in a patch of the QC, and all the states of each bulk band are given by resonances between all the replicas of this local pattern [66]. Interestingly, for a given bulk band, these local patches are arranged in an inflated QC [68]. Hence, we can think of each bulk band as an "effective photonic QC lattice," in the sense that each local patch acts as an "effective waveguide" (localizing the wave function), which is arranged in an inflated quasicrystalline pattern. The fact that each bulk band of the QC forms a new photonic QC manifests the fractal (self-similar) structure of the energy spectrum of the QC. Of course, an "effective QC lattice" given by a bulk band will not have the exact same tiling or the same nearest-neighbor couplings, and therefore its spectrum will not be a shrunken replica of the original spectrum. However, the "effective QC" maintains all the properties that come from its quasicrystal nature. This is the interpretation of self-similarity in the context of quasicrystals.

Here, we show that our photonic lattice displays these quasicrystal properties. Figure 2a shows the energy spectrum of the system [69] for $A_0 = 1.75$ using a periodic Penrose approximant. We can see that the spectrum has five main bulk bands that we label $-t, -d, s, +d, +t$. For these parameters, the dominant coupling of the effective Hamiltonian is the one associated with the waveguides at the minimum distance, i.e., the waveguides at the short diagonal of a thin rhombus, depicted as $d_0$ in Fig. 1d. As such, the $\{-d, +d\}$ bands are associated with symmetric or antisymmetric localization of the wave function in an isolated thin rhombus or "dimer" (the local pattern), while the $\{-t, +t\}$ bands are associated with coupled thin rhombuses or "trimers." Such local patterns are depicted in Fig. 2(c). This is in complete correspondence with Refs. [60,70], which study theoretically and experimentally the energy landscape of a nontopological realization of a Penrose lattice. If we zoom in on the band $-d$ as in Fig. 2(b), we can see the self-similar structure of the spectrum, in the sense that this band has several gaps that give rise to subbands within it. Now, since the band $-d$ is associated with isolated thin rhombuses, its subbands must be associated with combinations of them; specifically, isolated thin rhombuses form local patterns of three or five, as depicted in Fig. 2(c). For example, the $\{-3, -1, +1, +3\}$ subbands are associated with five isolated thin rhombuses, and the wave function is localized in such a

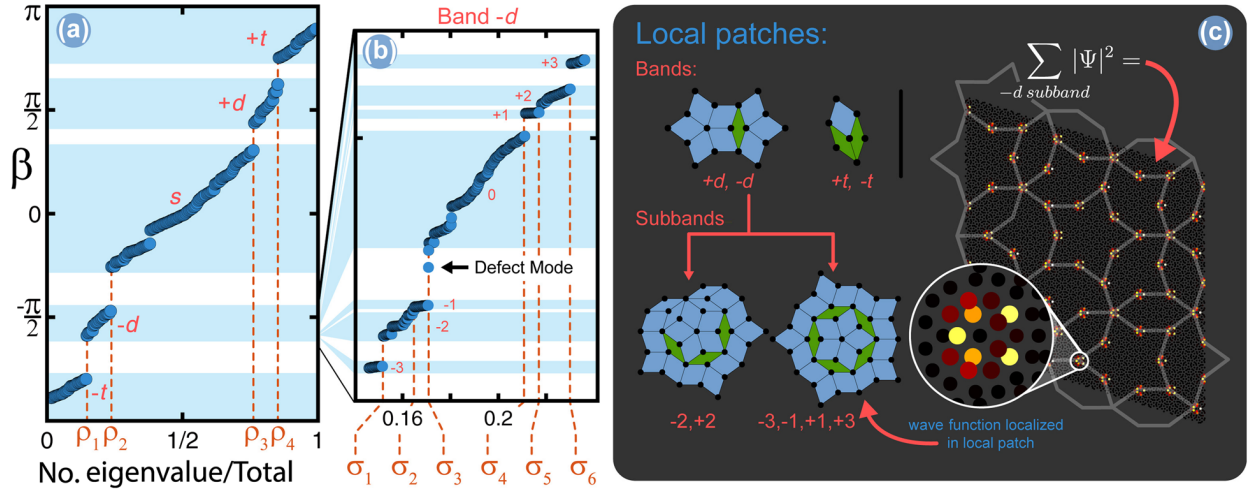

FIG. 2 (a) Quasienergy spectrum of a periodic Penrose tiling approximant containing 9349 vertices. $\rho_i$ mark the positions of the main gaps labeled (from the highest downwards) as $+t$, $+d$, $s$, $-d$, and $-t$. (b) Zoomed-in section of the quasienergy spectrum of quasiband $-d$. $\sigma_i$ mark the positions of the "subgaps" of the $-d$ band. (c) Left: Local patches associated with the $(+d, -d)$ bands and the $(+t, -t)$ bands, and local patches associated with the $-d$ band, giving rise to several subbands. Right: Sum of the modal intensities of the modes associated with the $-3$ subband.

local patch; see Fig. 2(c). As mentioned above, these local patches are arranged in an "inflated" quasicrystal tiling. To illustrate this, Fig. 2(c) shows the intensity sum of all the eigenfunctions for the $-3$ subband of the $-d$ band where we superimpose the inflated quasicrystal tiling. Interestingly, this gives a type of Penrose tiling that is not the original rhombus Penrose tiling we started with, but a Penrose tiling involving three prototiles: the star, the boat, and the cigar, defined in Ref. [70].

One of the most important spectral features of a QC is the "gap labeling" theorem [71], which shows that it is possible to identify, by a pair of integers, the position of the band gaps in the integrated density of states [72]. By studying the local patches of each band and its relative densities in an infinite tiling, we identify the position in integrated density of states of the four main band gaps: $\{\rho_1, \rho_2, \rho_3, \rho_4\}$. These gap locations are the same as the ones in the nontopological QCs studied in Refs. [60,70]. We also identify the positions of the six subband gaps of the second main gap, $\{\sigma_1, \sigma_2, \sigma_3, \sigma_4, \sigma_5, \sigma_6\}$. The positions of the band gaps are shown in Figs. 2(a) and 2(b), and they are given by $\rho_i = m_i + n_i \tau$, with $m_i = \{5, -3, 4, -4\}$, $n_i = \{-3, 2, -2, 3\}$, and $\sigma_i = (p_i + q_i \tau)/5$, with $p_i = \{148, -127, -4, 14, 137, -138\}$, $q_i = \{-91, 79, 3, -8, -84, 86\}$, and $\tau = (1 + \sqrt{5})/2$ is the golden ratio. As can be seen from Figs. 2(a) and 2(b), they are in excellent agreement with the numerically calculated spectrum.

Finally, to show the self-similar structure of the eigenstates, in Fig. 3 we show one bulk state for each of the main bulk bands of our photonic QC. Here, it is easy to see how the bulk states are formed by replicas of a same local pattern with different amplitudes, i.e., they are self-similar, and how they do not extend across the whole lattice.

Having confirmed the quasicrystalline properties of our helical QC lattice, we now study its topological properties. To have a clear view of the energy "landscape" of the QC, we calculate the density of states as a function of the strength of the vector potential $A_0$ (this corresponds to increasing the radius of the helical waveguides), for a periodic approximant, as shown in Fig. 4(a). As we can see, the QC presents a rich structure of band gaps that open and close as we change $A_0$. After identifying the gaps, we need to find out if the gaps are trivial or topological; i.e., we need to calculate their topological indices. In periodic systems, the topological index, the Chern number, can be readily calculated from the bulk band structure, i.e., the spectral structure of the system. This is not possible in QCs since the lack of periodicity excludes the possibility of having a well-defined band structure.

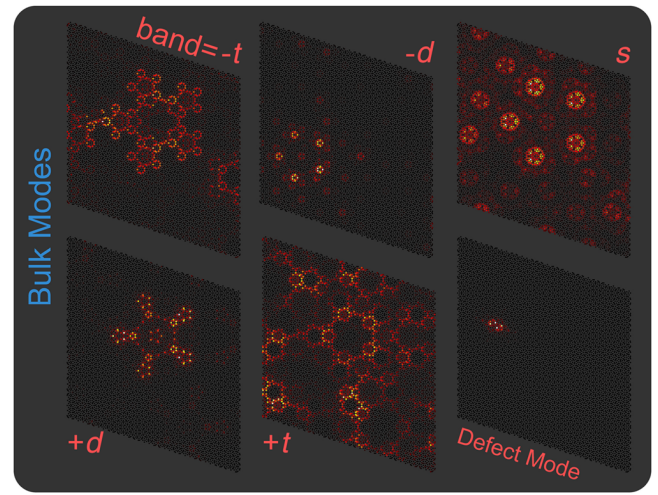

FIG. 3. Bulk states associated with each of the main bands $\{-t, -d, s, +d, +t\}$ of Fig. 2. Note the self-similar structure of the modes; i.e., they are formed by the repetition of a same pattern yet they do not extend over the entire lattice. Bottom right: A defect mode localized at one of the two mismatches of the Penrose approximants.

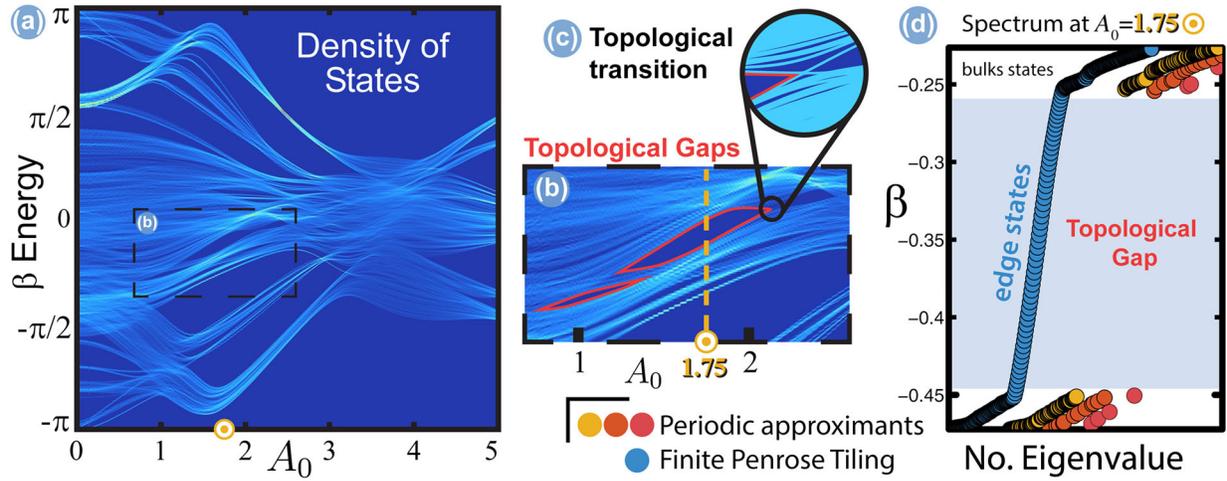

FIG. 4. (a) Density of states of the (periodic) Penrose tiling approximant, as a function of the strength of the vector potential $A_0$. (b) Zoomed-in section of the region marked by the frame in (a). The gaps enclosed by the red line are topological. (c) Zoomed-in section marked by the circle in (b), showing the transition from a topological gap to a nontopological (trivial) gap. As $A_0$ is increased, the topological gap closes and a trivial gap opens up. (d) Energy spectrum around the topological gap for periodic approximants with increasing system size (red, orange, and yellow) and for a finite Penrose tiling (blue). The approximants are periodic; hence, they have a full band gap. On the other hand, the finite Penrose tiling has edge states whose energies reside in the topological gap.

However, we can use the Bott index, a topological invariant especially useful for disordered systems [53,73]. The advantage of the Bott index is that it relies only on the spatial structure of the Hamiltonian. It measures the commutativity of the band-projected position operators given by $U = P\exp(i2\pi\hat{x})P$ and $V = P\exp(i2\pi\hat{y})P$, where $P$ is the projector onto the states with energy below a given band gap and $\{\hat{x}, \hat{y}\}$ are the normalized coordinate operators. The Bott index of a band gap at quasienergy $\beta$ is given by $C_B(\beta) = \text{Im}\{\text{Tr}[\log(VUV^\dagger U^\dagger)]\}/2\pi$ [53]. Physically speaking, the importance of the Bott index is that when it is nontrivial [$C_B(\beta) \neq 0$], it is not possible to find a complete and orthogonal basis of localized functions (Wannier functions) spanning the occupied states [53]. The inability to find localized Wannier functions implies that it is impossible to connect the Hamiltonian to a trivial one [53]. As such, a nontrivial Bott index implies the presence of a topological band gap. In fact, the Bott index can be shown to be equivalent to the Kubo formula for the Hall conductivity [53,76]. The Bott index $C_B(\beta)$, at a particular band gap (label by its quasienergy $\beta$), is equal to the net number of edge states at the band gap. In this way, the Chern number of a quasienergy band is simply the difference between the Bott indices at the band edges. We calculate the Bott index for all the band gaps visible in Fig. 4(a), and we find two large topological gaps—as depicted in Fig. 4(b). These band gaps have Bott index equal to $C_B = +1$, which indicates the presence of one edge state in the finite system. Actually, in Fig. 4(c) around $A_0 = 2.25$ it is possible to see how, by increasing $A_0$, the topological band gap closes and a trivial band gap opens up. This gap closing and reopening, as seen in Figs. 4(b) and 4(c) is a signature of the transition from the trivial to the topological phase. This type of crossing is reminiscent of a band inversion in periodic topological insulators.

There are two common distinct approaches to investigating topological phases: one is via their topological index and the other is via transport in the system. The approaches are related by the bulk-edge correspondence. Thus far, we have proved that it is possible to have a topological phase in a QC by showing it has nontrivial Bott index. Next, we study this topological phase via transport. In Fig. 4(d) we show the quasienergy spectrum around the topological gap of the photonic quasicrystal ($A_0 = 1.75$). The red, orange, and yellow dots indicate the spectrum using a periodic Penrose approximant with increasing system size, showing that—as the system size is increased—the band gap remains stable. However, since we are using approximants the system is periodic, a torus, and therefore one would not expect to have boundary defect states inside any of the gaps. Nevertheless, in some gaps there are one or two states inside the band gaps, as depicted in Fig. 2(b), whose corresponding "defect" mode is shown in Fig. 3. Interestingly, these states are due to the two mismatches of the periodic Penrose approximants and are localized on them; these defect states are not present in the perfect Penrose tiling. Now, when we calculate the quasienergy spectrum for a finite perfect Penrose lattice [blue dots in Fig. 4(d)], gapless edge states fill the same gap we identify to be topological using the Bott index. Figure 5 shows an edge state in this topological gap for various Penrose tiling with increasing system size. We can see that the topological edge states are localized on the edge of the quasicrystal and, therefore, they are extended along the system's perimeter, contrary to the general states of a quasicrystal system which have a power-law decay.

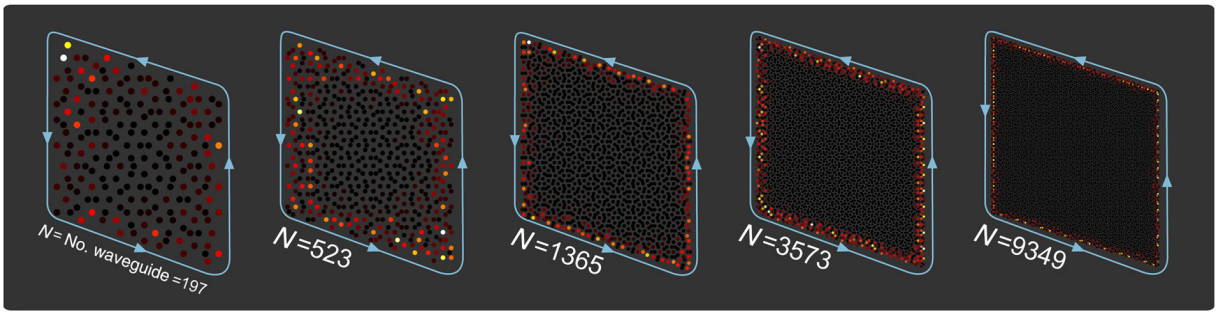

FIG. 5. Topological edge states associated with the topological band gap depicted in Fig. 4(d), for increasing size of the quasicrystal. The energy flow is always unidirectional as indicated by the blue lines.

To go further in showing the topological nature of the gapless edge states, we calculate the eigenmodes of photonic QC for different shapes of the bulk. In Fig. 6 we show the corresponding topological edge state for the different shapes. As expected for a topological system, the edge state is confined around the perimeter of the system irrespective of its shape; for example, a circle as in Fig. 6(a) or a random shape as in Fig. 6(b). Also, the edge state localizes around a hole in the system as depicted in Fig. 6(c). Even a single missing site (a single waveguide) or a mismatch (like the ones in the approximants) in the QC is enough to localize an edge state, creating a chiral in-gap bound state around such a defect [77,78], as shown in Fig. 6(d) for a missing waveguide. Here, it is important to point out that this state is a chiral bound state and not just a trivial defect state caused by the missing site. That is, boundaries usually create defect states inside trivial gaps, but these states are always localized and not extended around the whole perimeter as the topological edge states are. A missing site could also create a localized state residing inside a trivial gap, and attached to the site; however, the chiral in-gap bound state in Fig. 6(d) is localized "around" the missing waveguide, and one can check that it indeed has energy flux around it.

Now that we have shown the existence of gapless edge states in the topological gaps of our photonic QC, we study their dynamics, i.e., their unidirectional transport. We begin by creating a wave packet localized at the edge of the system by a superposition of gapless edge modes associated with the topological gap, and simulate their full evolution numerically. We do that by solving Eq. (1) numerically for continuous $z$ ("time") (instead of solving just for Floquet intervals). Figure 7 shows the propagation ("evolution") of a wave packet localized at the edge of the quasicrystal, for different bulk shapes. We can see clearly how the edge wave packet moves in one-way direction around the edge of the QC and how it is immune to backscattering from corners [Fig. 7(a)] or from the curvature of the edge [Fig. 7(b)]. Also note that diffraction broadening of the edge wave packet is very weak as it goes around the QC, similar to the evolution of topological edge states of periodic systems, which have an almost linear dispersion relation and therefore very weak diffraction. Since this edge wave packet lives in a topological gap with Bott index $C_B = +1$, it moves counterclockwise around the outer edge of the quasicrystal, as in Figs. 7(a) and 7(b). Interestingly, this topological edge state is moving clockwise if it is localized in a hole (or a defect) inside the QC, as in Fig. 7(c).

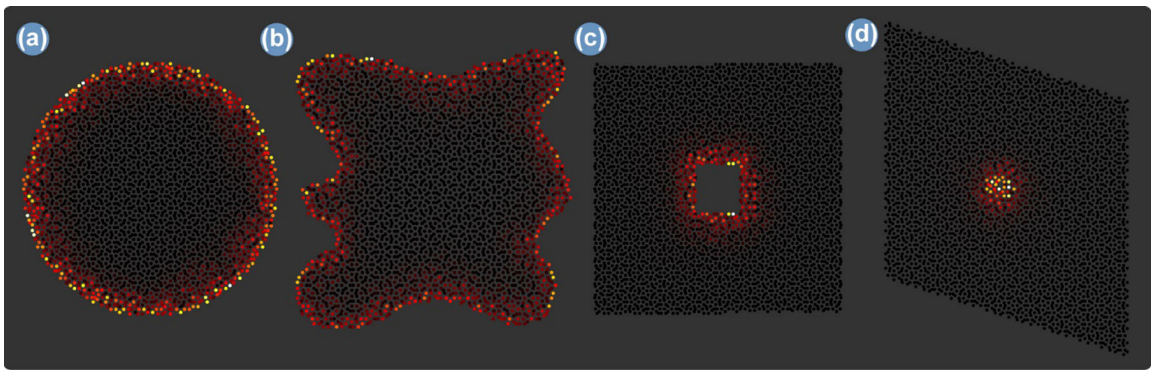

FIG. 6. Topological edge states associated with the topological gap shown in Fig. 4(d), for different shapes of the finite quasicrystal (a) circle, (b) arbitrary shape, (c) hole, and (d) a bulk defect (a missing waveguide in the bulk of the quasicrystal).

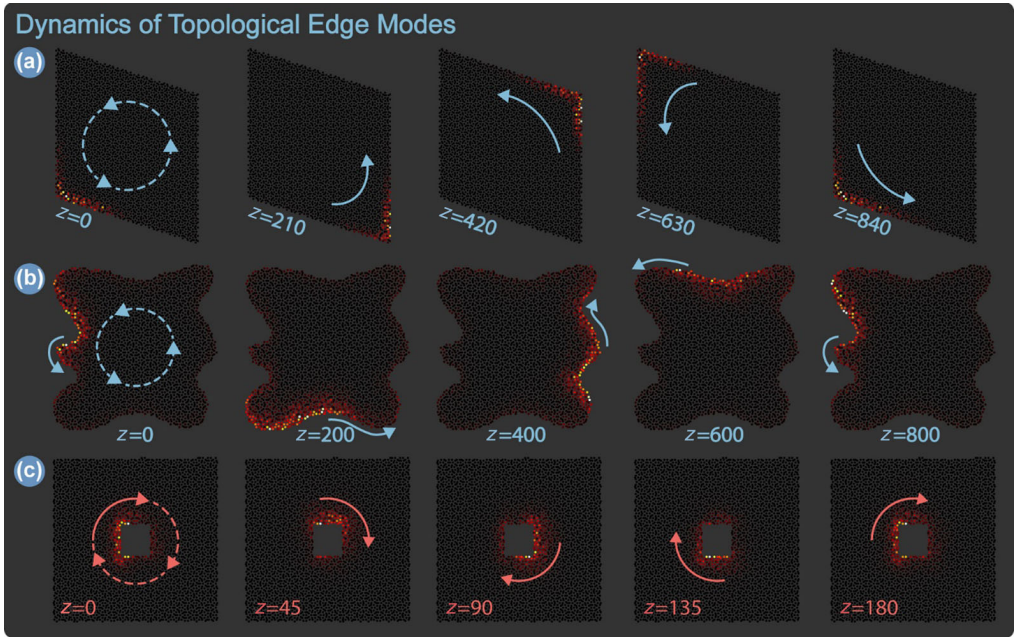

FIG. 7. Propagation dynamics of a topological edge wave packet moving along the edges of the quasicrystalline lattice (with $A_0 = 1.75$), for (a) square and (b) arbitrary-shaped lattices, and (c) for a square hole in the lattice. Here, $z$ denotes the propagation distance in units of the period of the periodic modulation. The propagation is always unidirectional, with no reflections or scattering into the bulk, even for sharp edges and arbitrary shapes. Because the Chern number is $+1$, the propagation is counterclockwise around the outer edge, as in (a) and (b), but clockwise in a hole inside the quasicrystal, as in (c).

Finally, we discuss the fractal topological properties of the energy spectrum of our photonic quasicrystal of helical waveguides. When the system size is increased, the major band gaps remain stable, but the system displays more and more minigaps (Fig. 8). When the QC lattice is modulated periodically (that is, the waveguides are made helical, i.e., $A_0 \neq 0$), we find that some of these new minigaps are topological. For example, Fig. 8 shows the energy spectrum of the photonic QC ($A_0 = 3$) and its topological band gaps, which have $C_B = -1$, for a Penrose tiling with 15466 vertices. There, it is possible to see the rich fractal structure of these topological minigaps; i.e., as one zooms closer and closer into the spectrum, it is possible to find more and more topological minigaps with nontrivial Bott index. One can also see how the gapless edge states fill the topological gaps (marked in orange in Fig. 8). These states will not be present if one uses a periodic Penrose approximant (on the torus), indicating that these are true topological band gaps, as described above. On the right-hand side of Fig. 8, we show one edge state associated with each of the topological gaps depicted in the spectrum (left-hand side of Fig. 8). Notice that the penetration depth of the edge state into the bulk increases as the size of the topological minigap decreases. For example, in Fig. 8, the topological edge state "$g$" resides in a small gap, and, therefore, it has more penetration depth into the bulk than the edge state "$h$," which resides in a big gap. This is why the presence of a topological band gap in the quasicrystal can be determined only by increasing the system size. The reason is that a very small band gap will possess edge states with large penetration depths into the bulk, necessarily requiring system sizes much larger than their decay lengths. This is in contrast to a periodic structure in which topological band gaps can be extracted by calculating topological invariants associated with the bulk band structure. Also, it is interesting to note that, although the topological edge states are extended along the edge, when they penetrate into the bulk they concentrate in the local patches that form the subbands near the corresponding gap. This effect is easy to see in the edge states of the smaller minigaps, for example, Figs. 8(a), 8(f), and 8(g). This is because the smaller the topological gap, the bigger the local patches of the subbands. *All of these features prove that what we present here is an entirely new phenomenon in which the fractal-like spectrum of a QC is endowed with topological band gaps.* In this way, the topological structure emerges as a function of the system size, contrary to periodic systems where the topological phase is completely characterized by the unit cell. We postulate that in our photonic QC an infinite number of topological gaps will open as the system goes to infinite size.

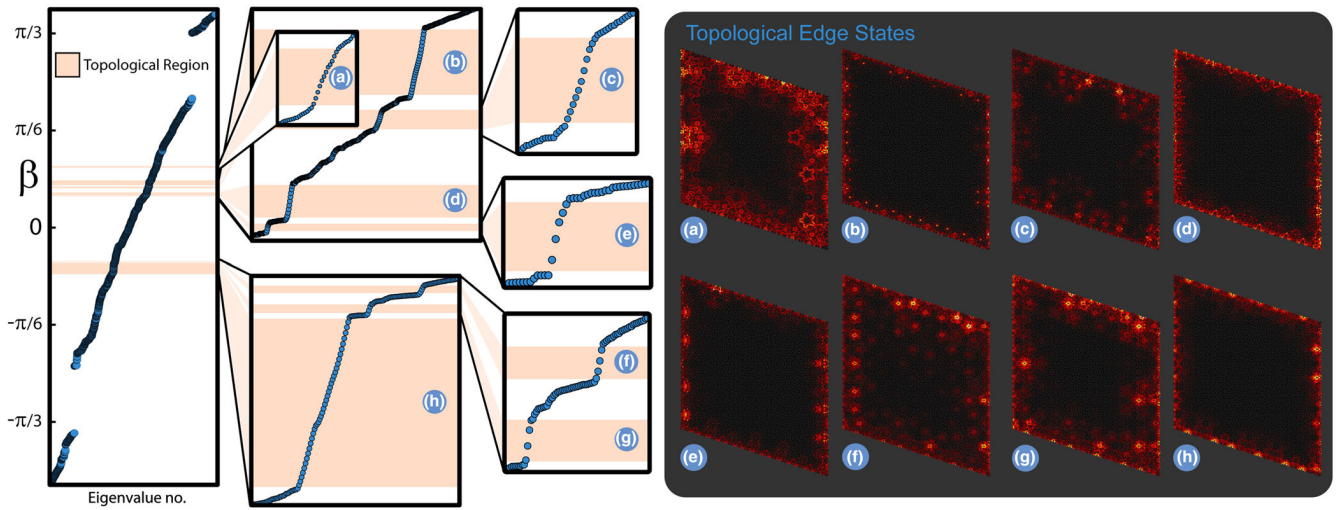

FIG. 8. The topological features occur for several levels of resolution of the energy spectrum. The left-hand panels show three generations of the fractal energy spectrum (with $A_0 = 3$ and 15466 vertices), where the pink regions mark the topological band gaps. The right-hand panel shows the topological edge states associated with the second and third "generations" of the fractal spectrum of a finite topological quasicrystal, where (a)–(h) associate each edge state to each of the gaps shown in the left-hand panel.

Before closing, we highlight the difference between the topological QC generated by an external magnetic field, proposed in Ref. [51], and the Floquet topological QC described here. The first is clearly the equivalent of the quantum Hall effect [5], whereas the latter is reminiscent of the quantum anomalous Hall effect (known as the Haldane model) [37], both implemented on a quasicrystalline lattice. The difference between the two is profound. In the case of Ref. [51], it is expected that in the presence of a magnetic field *any* lattice structure will have topological edge states, since they are present in the continuum (effective mass) limit. By contrast, without any magnetic field, only particular lattices (for example, those that have Dirac points) can have topological edge states; that is, the absence of a magnetic field makes it entirely unclear whether topological protection could arise. For quasicrystals, where Dirac points cannot be defined, it is doubly unclear whether topological gaps could be present in the absence of a magnetic field. Moreover, the spectrum in Ref. [51] has a Hofstadter-butterfly-like structure, which is a fractal energy spectrum that originates from the presence of a uniform magnetic field. In this case, the fractal spectrum of the Hofstadter butterfly exists whether the lattice is periodic or quasiperiodic, since there are two competing intrinsic length scales (the unit cell size of the underlying lattice and the magnetic unit cell), which may or may not be incommensurate. In contrast, in the work described here, the fractal nature of the spectrum originates solely from the underlying quasicrystal lattice.

In conclusion, we show that it is possible to have a topological phase in two-dimensional photonic QC with zero magnetic field. This topological Floquet photonic QC exhibits gapless unidirectional edge states that are extended along the system's perimeter, contrary to the general states of a QC system, which are critical, characterized by power-law decay. We have demonstrated the existence of this topological phase both by using a topological index (Bott index) and also by studying the dynamics of the gapless edge states. We have found an entirely new phenomenon in which the spectrum exhibits a rich fractal (self-similar) structure of topological minigaps. Hence, through the modulation, the fractal-like spectrum of a Floquet QC acquires topological bands, and the topological structure emerges as a function of the system size. This article shows that not only periodic systems with continuum spectra [10,30,37], or disordered systems with discrete spectra [36,38], can have topological properties with zero magnetic field, but also QC systems with singular continuum spectra. Our findings on this new topological phase in a QC raise many interesting questions for future work. For example, every known topological insulator seems to display extended bulk states that are robust against disorder [74,79]. Therefore, we can ask, are there extended bulk states in topological QCs? We conjecture that there are, but instead of being extended around the whole lattice, these bulk states extend thought the inflated lattices formed by the local patches that characterize that bulk band. In this way, our work lays the groundwork for further studies and we expect this new topological quasicrystal phase to be universal and applicable to a variety of quasicrystalline systems.

This work was supported by the Consejo Nacional de Ciencia y Tecnología, México. M. C. R. acknowledges the support of the National Science Foundation under Grant No. ECCS-1509546. This research was supported by the Transformative Science Program of the Binational USA-Israel Science Foundation (BSF) and by the Israeli ICore Excellence Center "Circle of Light." The research leading to these results has received funding from the European Union's–Seventh Framework Programme (FP7/2007-2013) under Grant Agreement No. 629114 MC–Structured Light.